\documentclass[runningheads,a4paper]{llncs}

\usepackage{amssymb}
\usepackage{graphicx}

\usepackage{url}
\urldef{\mailsa}\path|{c.jacobs10, a.avdis, simon.mouradian06, m.d.piggott}@imperial.ac.uk|
\newcommand{\keywords}[1]{\par\addvspace\baselineskip
\noindent\keywordname\enspace\ignorespaces#1}

\begin{document}

\mainmatter  

\title{Integrating Research Data Management into Geographical Information Systems}

\titlerunning{Data Management in Geographical Information Systems}
\authorrunning{Data Management in Geographical Information Systems} 

\author{Christian T. Jacobs \and Alexandros Avdis\and\\ Simon L. Mouradian\and Matthew D. Piggott}

\institute{Department of Earth Science and Engineering, South Kensington Campus,\\
Imperial College London, London SW7 2AZ, United Kingdom\\
\mailsa\\
\url{http://www.imperial.ac.uk/engineering/departments/earth-science}}

\maketitle

\begin{abstract}
Ocean modelling requires the production of high-fidelity computational meshes upon which to solve the equations of motion. The production of such meshes by hand is often infeasible, considering the complexity of the bathymetry and coastlines. The use of Geographical Information Systems (GIS) is therefore a key component to discretising the region of interest and producing a mesh appropriate to resolve the dynamics. However, all data associated with the production of a mesh must be provided in order to contribute to the overall recomputability of the subsequent simulation. This work presents the integration of research data management in QMesh, a tool for generating meshes using GIS. The tool uses the PyRDM library to provide a quick and easy way for scientists to publish meshes, and all data required to regenerate them, to persistent online repositories. These repositories are assigned unique identifiers to enable proper citation of the meshes in journal articles.

\end{abstract}

\section{Introduction}\label{sect:introduction}
\begin{figure}
   \centering\includegraphics[width=0.6\columnwidth]{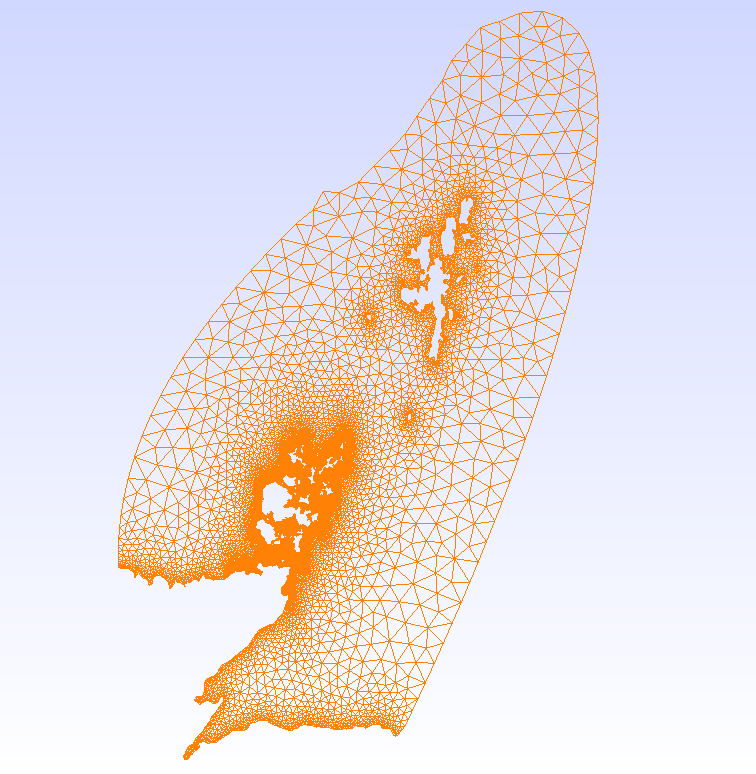}
   \caption{An example of an unstructured computational mesh which discretises the marine area around the North-East coast of Scotland. The resolution is highest around the Scottish coastline and around the Orkney and Shetland Isles.}
   \label{fig:mesh}
\end{figure}

Computer simulations of ocean dynamics are becoming ever more important to predict the effects of global-scale hazards such as tsunamis \cite{Hill_etal_2014}, the influence of marine renewable energy turbines on sediment transport \cite{MartinShort_etal_2015}, and the dispersal range of nuclear contaminants \cite{Choi_etal_2013}, to name just a few applications. The underlying numerical model behind such simulations often requires a mesh upon which the equations describing the flow dynamics are solved, thereby transitioning from a continuous description of the region of interest (also known as the domain) to a discrete one. An example focussing on the area around the Orkney and Shetland Isles is shown in Figure \ref{fig:mesh}. A mesh for ocean simulations must be of high enough quality to resolve the intricate coastlines and bathymetry \cite{Gorman_etal_2008}. However, creating such a mesh manually is infeasible for large-scale, high-resolution simulations.

Geographical Information Systems (GIS) offer an effective way of processing bathymetry and coastline data to create a geometry with which to work \cite{Li_2000}. A method of producing a computational mesh from this geometry is then required to perform a simulation on it. QMesh \cite{Avdis_etal_InPreparation} is a software package currently being developed at Imperial College London for this purpose. QMesh reads in a geometry defined in the QGIS Geographical Information System software \cite{QGIS_software}, and then converts the geometry into a readable format for the Gmsh mesh generation software \cite{GeuzaineRemacle_2009}, which in turn generates the mesh to provide a discrete representation of the domain. Ocean simulations may then be performed with a computational fluid dynamics package.

Publications that are dependant on numerical simulation often provide details of the simulation setups to improve reproducibility and indeed recomputability. However, while a description of the domain may also be given, the mesh that discretises this domain is rarely provided as a supplementary material. This lack of data availability has also been highlighted in many other areas of science \cite{Whitlock_etal_2010}, \cite{Alsheikh-Ali_etal_2011}, \cite{Vines_etal_2013}. Furthermore, citations to the software used to produce the mesh typically only refer to a generic user manual and contain no information about which version was used. For the purpose of recomputability and reproducibility, it is crucial that researchers provide \textit{all} the data files, as well as the precise version of the software's source code used to produce the output in the first place \cite{deLeeuw_2001}, \cite{BuckheitDonoho_1995}. In the case of this work, the input data is the geographical information defining the domain, the output data is the computational mesh, and the software is QMesh (and its dependencies).

Despite the need for a more open research environment where software and datasets are shared freely, the level of motivation amongst researchers to do this is generally quite low. This is in part due to the extra effort and time required to gather and publish the data \cite{LeVeque_etal_2012}, whilst typically gaining little from the process. To encourage the sharing of data and improve its reproducibility and recomputability, it is therefore important to make the publication process more straight forward and swift. This can be effected by the development of research data management tools that readily capture the datasets involved and information about the software being used \cite{Stodden_etal_2013}, \cite{LeVeque_etal_2012}.

This paper describes the integration of a research data management tool, which uses the PyRDM library \cite{Jacobs_etal_2014a}, into the QMesh software. The tool automates the publication of the QMesh source code, as well as the input and output data for a specified QGIS project, to online, citable and persistent repositories such as those provided by Figshare (figshare.com), Zenodo (zenodo.org) and DSpace-based (dspace.org) hosting services. The tool has both a command line and a graphical user interface, and allows users to publish the software and data at the `push of a button', thereby facilitating sharing and a more open research environment. In contrast to other software tools that also facilitate the publication of code and datasets, such as Fidgit \cite{Smith_2013}, rfigshare \cite{Boettiger_etal_2014}, and dvn \cite{Leeper_2014}, the QMesh publishing tool incorporates application-specific knowledge to provide a greater amount of automation. For example, the tool is able to parse QGIS project files to automatically determine the relevant input data to publish, rather than the user having to specify the data files manually. Furthermore, this work represents a novel application of research data management and curation software within a GIS environment.

Section \ref{sect:integration} describes in greater detail the extensions made to the QMesh software to automate the publication process for the software itself, the input files (for a given QGIS project) and any output files (i.e. the computational mesh). Section \ref{sect:example} presents a realistic example of a scientific workflow involving production of a mesh of a UK coastal region. The data files are read in to QGIS and a mesh is produced. Both the QGIS data and mesh are subsequently published to an online repository provided by Figshare, and a DOI is assigned which can be used to properly cite the data in journal articles. Finally, some concluding remarks are made in Section \ref{sect:conclusions}.

\section{Integration with QMesh}\label{sect:integration}
QMesh features a command line interface (CLI), as well as a graphical user interface (GUI) via a QGIS plugin through which users can select relevant geometry objects and produce a mesh. The integration of research data management techniques into QMesh was achieved by adding a PyRDM-based publishing tool to both of these interfaces.

\begin{figure}
   \centering\includegraphics[width=0.75\columnwidth]{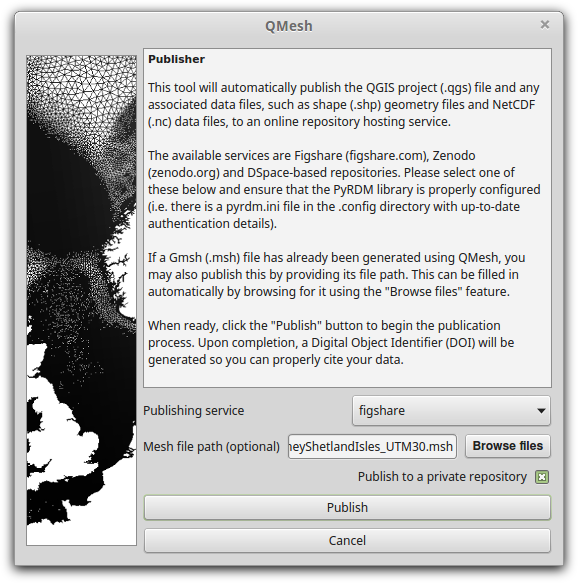}
   \caption{The QMesh publisher tool, which is part of the QMesh QGIS plugin. Users choose the online repository service that they wish to use; by default this is set to Figshare. In addition to the input data files associated with the QGIS project, users may also publish the output data file (i.e. the resulting computational mesh) produced by QMesh, if they so desire. By default, the publication is made public unless the user decides otherwise; in the case of private publication, a DOI is still assigned to the repository, but will not be made active/`live' until the repository is made public.}
   \label{fig:publish_dialog}
\end{figure}

The tool provides the option of publishing the QMesh software source code and data required to reproduce the mesh to separate online repositories. Users are presented with a simple interface and only have to provide a minimal amount of information; this is illustrated in Figure \ref{fig:publish_dialog}. The publication process itself is handled by the PyRDM library \cite{Jacobs_etal_2014a} which communicates with an online repository hosting service via its Application Programming Interface (API). The publication process results in a Digital Object Identifier (DOI) \cite{DavidsonDouglas_1998} being assigned to the repository, with which users can properly cite their research outputs.

\begin{figure}
   \centering\includegraphics[width=0.44\columnwidth]{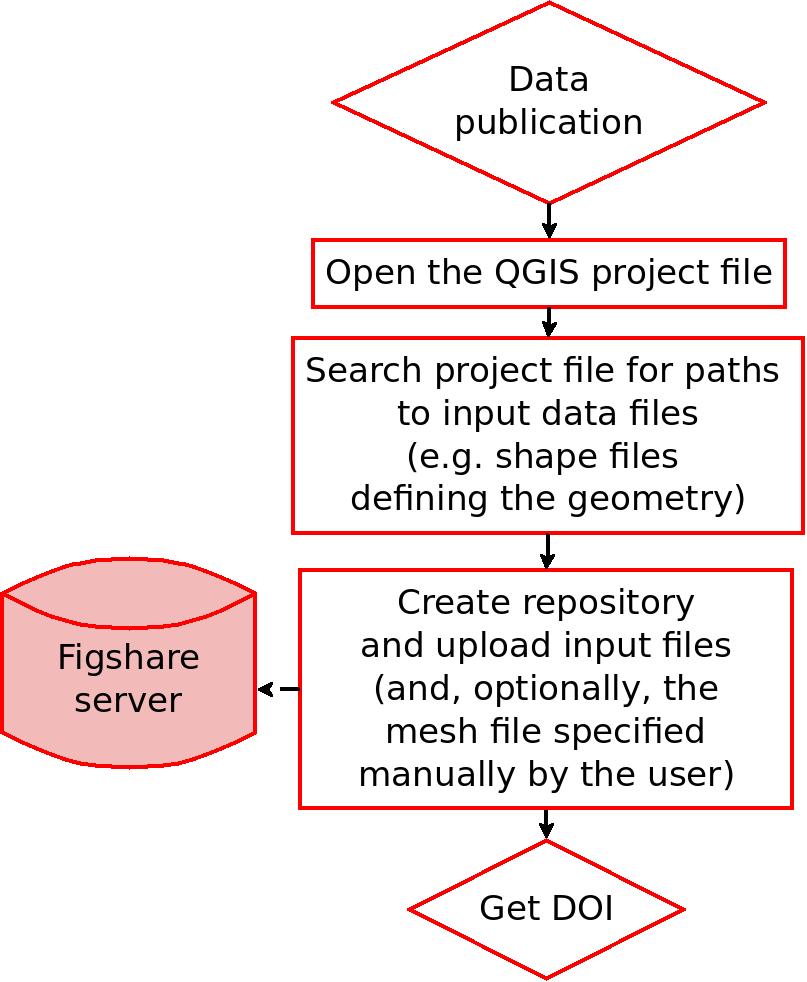}
   \hspace*{10mm}\centering\includegraphics[width=0.44\columnwidth]{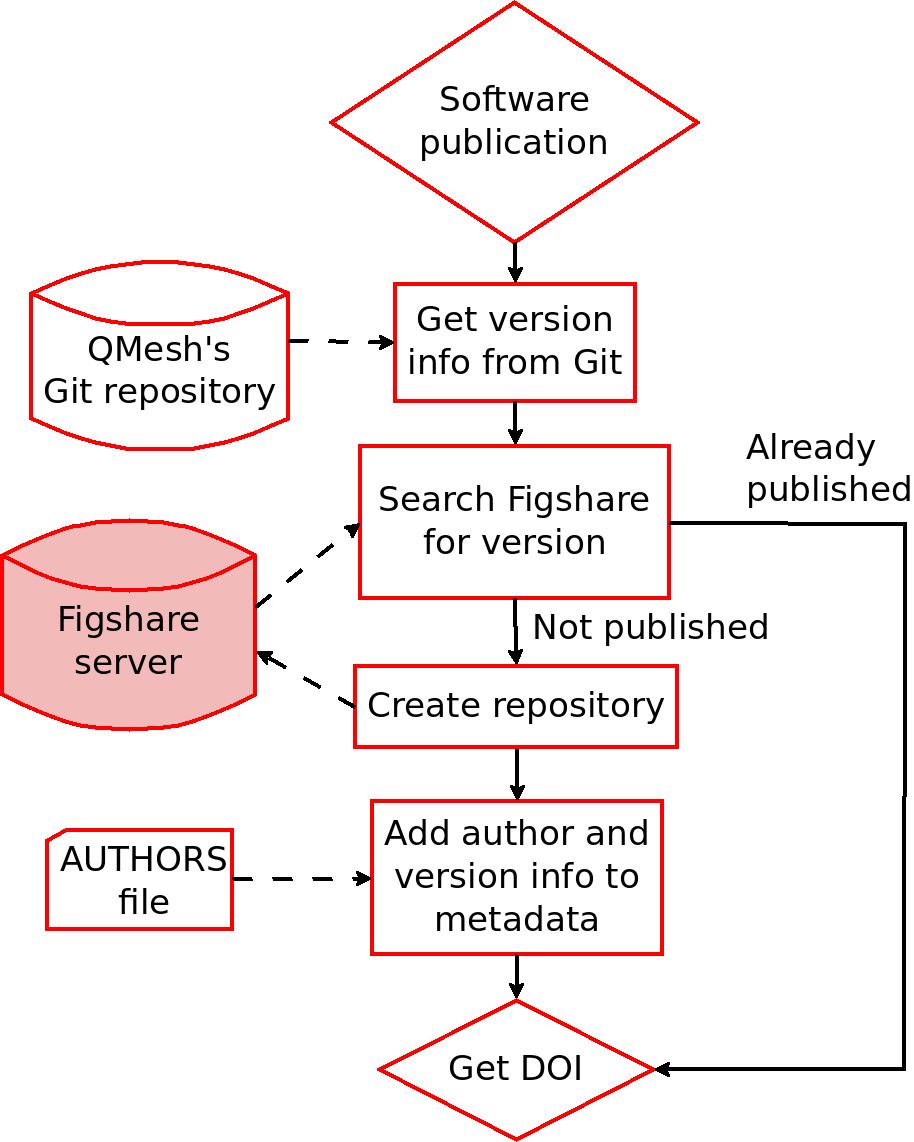}
   \caption{The processes behind publishing the QGIS data files (left) and QMesh software source code (right) to Figshare.}
   \label{fig:publishing_process}
\end{figure}

The publication of data is handled separately to the publication of the QMesh software. In the former case, when a suitable mesh has been produced and is ready to be published, users simply have to provide the QMesh publishing tool with the location of the QGIS project file on the computer's file system when using the CLI. When using the GUI, this location is provided automatically when the project is opened in QGIS. The tool then searches for the \texttt{<datasource>} tags in the XML-based project file to determine the location of all the files that the project comprises; these may include shape files that define various layers in the geometry, data files in NetCDF format \cite{RewDavis_1990} which define the bathymetry of the ocean, and a multitude of other data formats. Optionally, the location of the Gmsh mesh file may also be provided, thereby publishing the resultant output data along with the files required to produce it. The locations of all these data files, including the QGIS project file itself, are then provided to PyRDM which automatically creates a repository on the hosting service and uploads the files via the service's API. The service then returns a publication ID and a DOI, which is presented to the user for citation purposes. This process is illustrated in Figure \ref{fig:publishing_process}.

The publication of software involves a similar process, but can currently only be accomplished via the CLI. The user only has to provide the QMesh publishing tool with the location of the software's source code on the computer's file system. The PyRDM library then handles the rest; it determines the exact version of QMesh currently in use using the Git version control system (git-scm.com) \cite{Ram_2013}, and then checks to see whether that version has been published already\footnote{Repository searching is only available when using the Figshare repository service, due to API limitations explained later in Section \ref{sect:conclusions}. PyRDM will publish the software regardless of whether it has been published before when Zenodo or a DSpace-based service is chosen.}. If it has, PyRDM retrieves the existing DOI for re-use. If it has not, then PyRDM publishes the source code in a similar fashion to the case of publishing data, as shown in Figure \ref{fig:publishing_process}. Note that publications in journals would need to reference both the software repository's DOI and the data repository's DOI. There is currently no explicit link that is made between the software and data repositories, unless specified manually.

As demonstrated by Figure \ref{fig:publishing_process}, the QMesh publishing tool requires minimal user interaction and is largely automated by the PyRDM library. This is important for encouraging the sharing of software and data files, in order to achieve a more open research environment.

\section{Workflow Example}\label{sect:example}
\begin{figure}
   \centering\includegraphics[width=1\columnwidth]{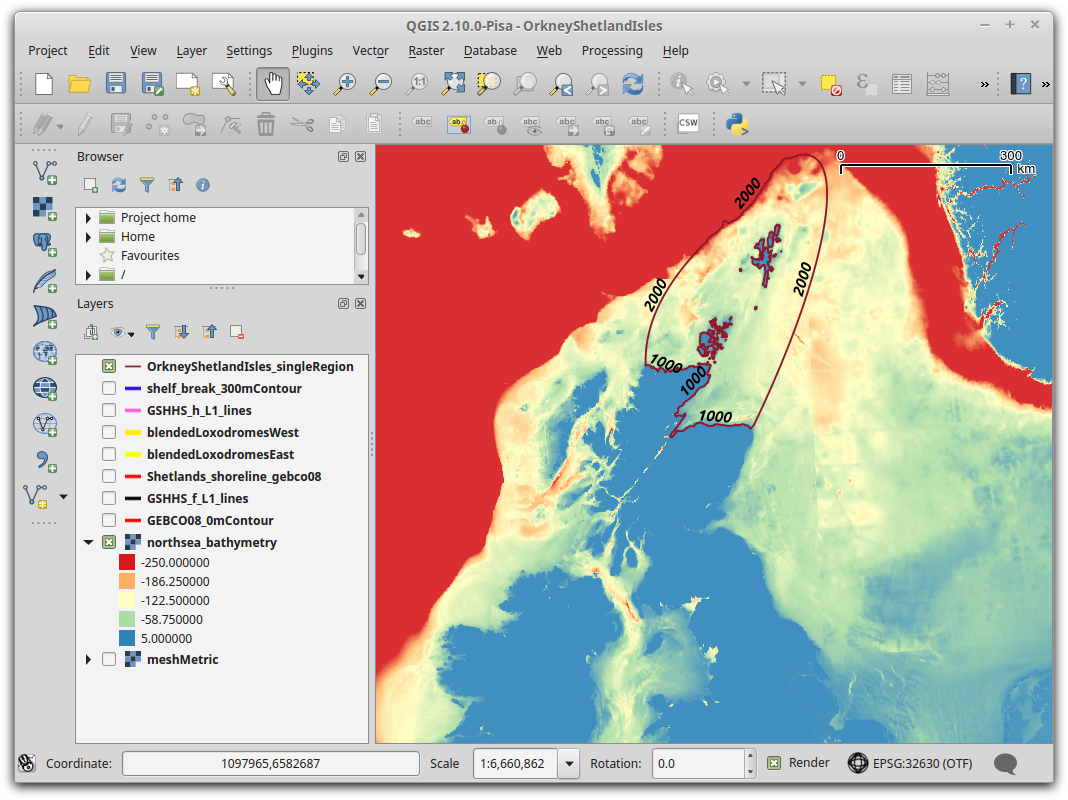}
   \caption{Screenshot of the UK region visualised in QGIS. The solid dark purple line defines the area that will be meshed (in this case it contains the Orkney and Shetland Isles). The different files that make up the layers of the geometry are specified in the column on the left-hand side.}
   \label{fig:qgis_geometry}
\end{figure}

To demonstrate an example of a scientific workflow involving mesh generation using GIS, the Orkney and Shetland Isles considered in \cite{Avdis_etal_InPreparation} and \cite{Avdis_etal_Accepted} are used. The researcher first has to describe the geography of the domain in QGIS and then decide on the area they wish to create a mesh for.  The QGIS project for the Orkney and Shetland Isles comprises a number of geometrical layers which define the coastlines (and potentially coastal engineering structures such as marine power turbines), in addition to a NetCDF file which defines the bathymetry of the ocean floor, and another NetCDF file which defines the desired resolution throughout the mesh. These files are shown in Figure \ref{fig:qgis_geometry} beside the area that will be meshed.

\begin{figure}
   \centering\includegraphics[width=1\columnwidth]{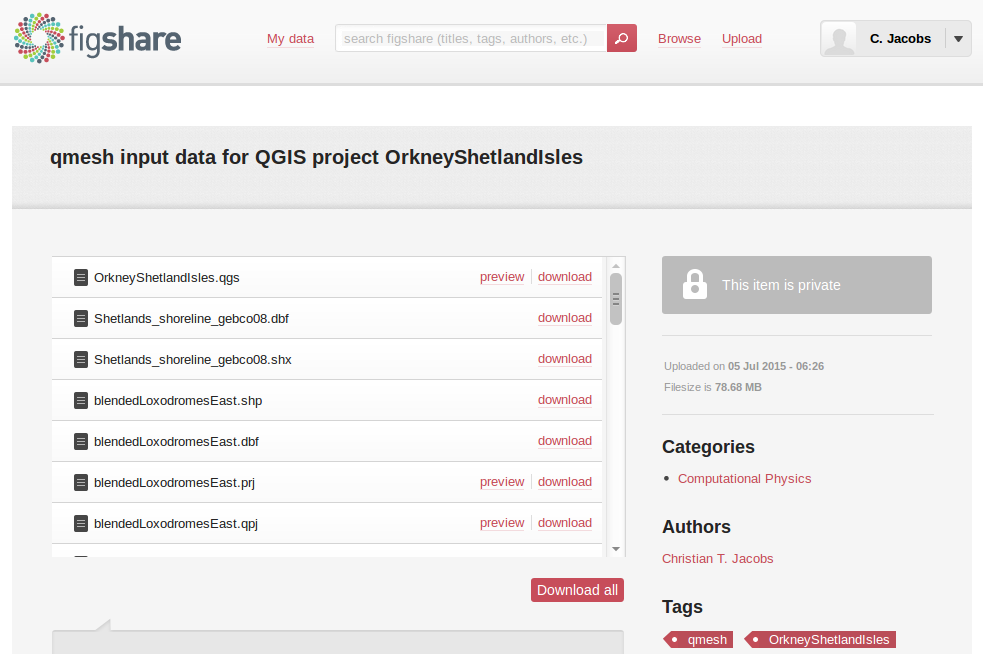}
   \caption{A screenshot of the resulting repository on the Figshare website, with the files readily available to download. The QMesh publishing tool automatically assigns a title and tags to the repository based on the QGIS project's name.}
   \label{fig:figshare_files}
\end{figure}

\begin{figure}
   \centering\includegraphics[width=0.75\columnwidth]{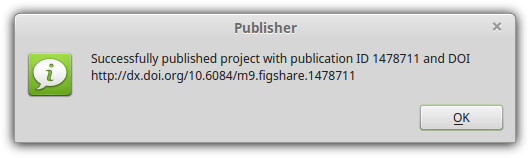}
   \caption{A Figshare publication ID and a DOI are assigned to each repository, and presented to the researcher once the publication process is complete.}
   \label{fig:doi}
\end{figure}

The mesh that QMesh produces for this domain (shown in Figure \ref{fig:mesh}) is then used by the researcher in their marine simulations. Once the researcher is ready to publish their results, they upload the data files associated with the production of the simulation's mesh to an online repository using the QMesh publishing tool shown in Figure \ref{fig:publish_dialog} (the CLI may also be used instead of the graphical interface). In this example, it uploads all the files previously mentioned to Figshare. Once uploaded, the files can be downloaded from the Figshare website (see Figure \ref{fig:figshare_files}) and a DOI is presented to the researcher to share with colleagues and for use in journal publications (see Figure \ref{fig:doi}).

The version of the QMesh software's source code that is used should also be published, in a separate repository to the data. However, it should be noted that publishing the QMesh source code may not be enough to reproduce the exact same mesh without also knowing the versions of its dependencies. For example, different versions of Gmsh may produce slightly different meshes as a result of algorithmic improvements within the software. It is therefore important that such information be recorded in some way to further improve the degree of reproducibility. For example, ideally Gmsh would also have a similar system for publishing the current version of its source code in use.

\section{Discussion and Conclusions}\label{sect:conclusions}
Throughout the production of the PyRDM-based publishing tool for QMesh, several issues were encountered which largely stemmed from a lack of standardisation and support in the repository hosting services' APIs. For example, in order for PyRDM to attribute authors to the software repository on Figshare, all authors of QMesh must provide their Figshare author IDs in the \texttt{AUTHORS} file that is part of the QMesh source code. Unfortunately, another different set of author IDs would need to be provided when using a different repository service such as Zenodo, which is inconvenient and requires all authors of QMesh to have accounts across all the supported services. A more standardised way of identifying and attributing authors to research software and data would be to use ORCID (orcid.org) researcher IDs. Figshare has recently added support for authenticating with ORCID IDs via its web interface \cite{Figshare_2013}, and it is hoped that ORCID authentication via the Figshare API will also be added for the benefit of PyRDM. Another example, this time involving lack of API support, is the current inability to search for an existing repository with the Zenodo API. Further developments are necessary in this area to enrich the publication process and improve automation.

The production of meshes can involve proprietary and/or private data which cannot be published openly, but at the same time sharing all research output is becoming a common requirement imposed by research funders. The QMesh publishing tool comes with the option of publishing the data to private repositories. However, with some services the private storage space is rather limited, and typically not large enough to store high quality mesh files for realistic ocean simulations. For example, the free private storage space offered by Figshare is 1 GB at the time of writing this paper, with a 250 MB individual file size limit\footnote{\texttt{http://figshare.com/pricing}}. Furthermore, only a maximum of 5 collaborators can be given access to a private repository. In contrast, the integration of Figshare for Institutions \cite{Figshare_2014} offers a more suitable platform for larger-scale research data management. This project enables researchers at an institution to publish to private repositories hosted in the cloud. This is considerably more sustainable for GIS projects and mesh generation that can involve very large file sizes, both public and private data, and collaboration amongst many researchers and research groups.

In conclusion, the integration of a publishing tool in a Geographical Information System has helped to mitigate one of the reasons why researchers tend not to publish their software and data; that is, it is time-consuming to do so with little reward. The new QMesh publishing tool makes publishing a computational mesh and associated data files easy and largely effortless through the addition of a significant amount of automation. Furthermore, the use of online repository services enable more formal citation of all research outputs through the use of DOIs. However, it is the responsibility of the scientific community to encourage and provide incentives for the openness and public availability of this software and data, in order to overcome the barrier of lack of motivation to publish.

\subsubsection*{Acknowledgments}
CTJ was funded by an internal grant entitled ``Research data management: Where software meets data'' from the Research Office at Imperial College London. Part of the work presented in this paper is based on work first presented in poster form at the International Digital Curation Conference (IDCC) in February 2015 \cite{Jacobs_etal_2015}, and in a PyRDM project report \cite{Jacobs_etal_2014b}. The authors would like to thank the two anonymous reviewers of this paper for their feedback.

\bibliographystyle{splncs03}
\bibliography{gis-rdm}

\end{document}